\documentclass[aps,pre,twocolumn,groupedaddress,floatfix]{revtex4-1}

\usepackage[english]{babel}
\usepackage[utf8]{inputenc}
\usepackage[T1]{fontenc}
\usepackage{xcolor}
\usepackage{graphicx}
\usepackage{psfrag}
\usepackage{amsmath}
\usepackage{bm}
\usepackage{bm}
\usepackage{hyperref}
\hypersetup{
    colorlinks=true,
    citecolor=blue,
    linkcolor=blue,
    urlcolor=blue,
}
\hypersetup{pdfstartview={FitH}}

\newcommand{\dd}{{\rm d}}

\allowdisplaybreaks

\begin{document} 
	
\title{Current-density relation in the exclusion process with dynamic obstacles}
	
\author{J. Szavits-Nossan$^1$, B. Waclaw$^{1,2}$}
	
\affiliation{$^1$School of Physics and Astronomy, University of Edinburgh, Peter Guthrie Tait Road, Edinburgh EH9 3FD, United Kingdom\\
$^2$Centre for Synthetic and Systems Biology, University of Edinburgh, Edinburgh, United Kingdom}
	
\begin{abstract}
We investigate the totally asymmetric simple exclusion process (TASEP) in the presence of obstacles that dynamically bind and unbind from the lattice. The model is motivated by biological processes such as transcription in the presence of DNA-binding proteins. Similar models have been studied before using the mean-field approximation, but the exact relation between the particle current and density remains elusive. Here, we first show using extensive Monte Carlo simulations that the current-density relation in this model assumes a quasi-parabolic form similar to that of the ordinary TASEP without obstacles. We then attempt to explain this relation using exact calculations in the limit of low and high density of particles. Our results suggest that the symmetric, quasi-parabolic current-density relation arises through a non-trivial cancellation of higher-order terms, similarly as in the standard TASEP.
\end{abstract}
	
\maketitle
	
\section{Introduction}
\label{introduction}

Totally asymmetric simple exclusion process (TASEP) is a paradigmatic model of non-equilibrium statistical mechanics. In its simplest version, particles enter a one-dimensional lattice of size $L$ from one boundary and exit through the other boundary. A particle jumps at a constant rate to the next lattice site, unless that site is blocked by another particle.

\begin{figure}[hb]
	\includegraphics[width=3.2in]{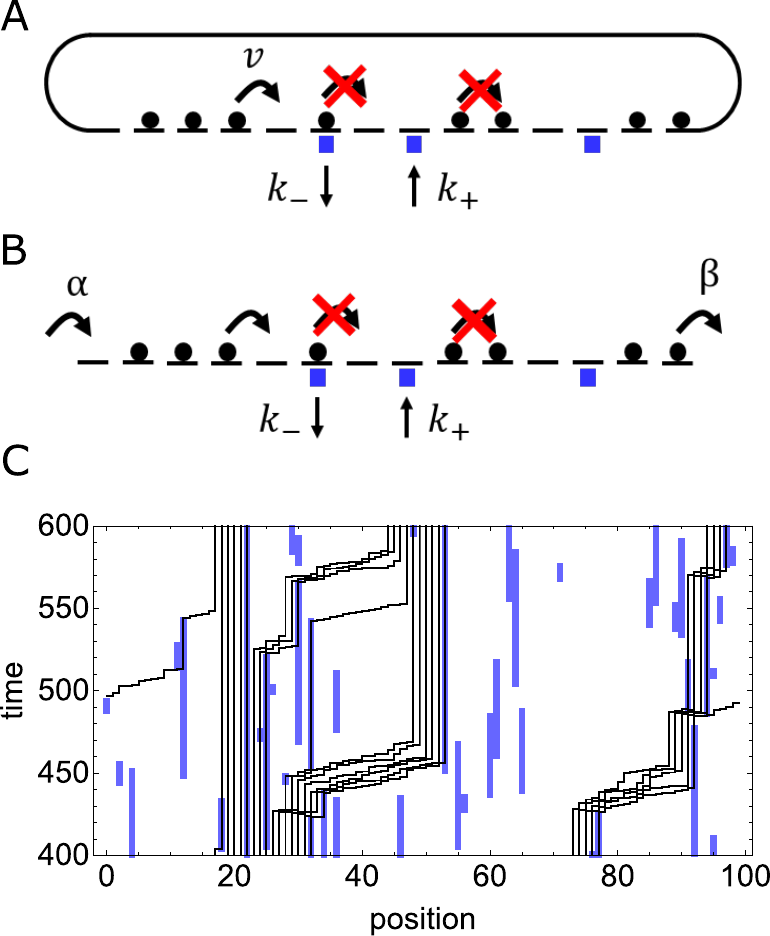}
	\caption{\label{fig1} The TASEP with dynamic obstacles. (A) The model with periodic boundary conditions. (B) The model with open boundaries. (C) A space-time plot of a simulation snapshot for the model with periodic boundaries, for $L=100, M=20, k_+=0.002, k_-=0.02$. Particles are black, obstacles are blue. Particles can be seen moving to the right and stopping when encountering obstacles. Obstacles appear at random sites and disappear after a while.}
\end{figure}

The TASEP was originally proposed to model the dynamics of ribosomes during mRNA translation \cite{MacDonald1968,MacDonald1969}, and it is still in much use for that purpose today \cite{Zia2011,vonderHaar2012,Garai2009a,Garai2009b, Ciandrini2013,Zur2016,Szavits2018a,Scott2019,ErdmannPham2020}. Other applications include enzyme kinetics \cite{Shapiro1982}, the dynamics of RNA polymerases during transcription \cite{Tripathi2008,Heberling2016,vandenBerg2017,Fernandes2019,CholewaWaclaw2019}, the movement of molecular motors \cite{Chowdhury2005,Miedema2017}, and modelling pedestrian and vehicular traffic \cite{Chowdhury2000,Schadschneider2010}.

From the theoretical point of view, the TASEP has been extensively studied as a model system for boundary-driven phase transitions \cite{Krug1991,Schmittmann1995}. The model with open boundary conditions is one of few models in nonequilibrium statistical mechanics  which can be solved exactly \cite{Derrida1993,Schutz1993}. Interestingly, bulk properties of the TASEP such as the particle current and density can also be explained using the mean-field approximation that ignores correlations between particles \cite{MacDonald1968,Derrida1992}. Other examples where the mean-field approximation has been applied successfully include extended particles (e.g. ribosomes that occupy $\approx 10$ lattice sites) \cite{Shaw2003}, particles that can attach and detach from the lattice (the Langmuir kinetics) \cite{Parmeggiani2003}, particles with site-dependent hopping rates \cite{Janowsky1992,Chou2004,Shaw2004,ErdmannPham2020} and particles with internal states \cite{Klumpp2008,Ciandrini2010,Wang2014}.

A remarkable property of the standard TASEP is a simple relation between the current $J$ of particles and particle density $\rho$ in the thermodynamic limit ($L\to\infty$):
\begin{equation}
    J = 4 J_{\rm max} \rho (1-\rho) , \label{eq:Ju}
\end{equation}
where $J_{\rm max}$ is the maximum current that occurs when $\rho=1/2$. Due to the particle-hole symmetry, this relation is symmetric to $\rho\leftrightarrow 1-\rho$. For the standard TASEP, $J_{\rm max}=v/4$, where $v$ is the particle hopping rate. We will refer to Eq.~(\ref{eq:Ju}) as the \emph{current-density relation}. This relation can be derived using either the mean-field approximation or the exact solution. 

Here we explore to what extent the above relation holds for the TASEP with dynamic ``defects'' (henceforth the acronym ddTASEP) that temporarily block or slow down the movement of particles (Fig. \ref{fig1}). Many variants of the TASEP with dynamic disorder exist in the literature, such as a single dynamical defect \cite{dong_inhomogeneous_2007,turci_transport_2013,sahoo_dynamic_2015}, annealed disorder \cite{das_particles_2000,barma_driven_2006} and the ``bus route" model \cite{bus-route}. Here, we focus on the case motivated by interactions between the RNA polymerase and DNA-binding proteins \cite{Ghosh2018,CholewaWaclaw2019}, in which particle-blocking obstacles hop in- and out of the lattice from an infinite reservoir. The lattice can therefore have an arbitrary number of obstacles between $0$ and $L$ and any site can be occupied by an obstacle.

In our previous paper \cite{Waclaw2019} we used computer simulations to show that the parabolic form of Eq. (\ref{eq:Ju}) holds approximately for the ddTASEP, whereby the effective particle speed $u=4J_{\rm max}$ was found to be smaller than the intrinsic speed $v$. We also applied a simple mean-field approximation to derive the current-density relation for high binding and unbinding rates. Our result broke down when the binding and unbinding rates were low. A similar failure of the mean-field approximation was previously reported in the TASEP with an isolated dynamic defect \cite{Turci2013}. 

In the present work, we would like to better understand the origin of this quasi-parabolic current-density relation in the ddTASEP. 
We use Monte Carlo computer simulations to show that the current-density relation is in fact not a parabola, but that deviations from the parabolic shape remain very small over a large range of binding and unbinding rates. We then consider the dynamics of a single particle on an infinite lattice and derive an expression for $u$ that agrees well with the results of computer simulations. Finally, we derive the current-density relation for the system with open boundary conditions using the power series method \cite{Szavits2013,Szavits2018b}, and show that particle-defect correlations dominate over particle-particle correlations in setting the maximum current.

\section{The model}
\label{model}

In contrast to \cite{Ghosh2018,CholewaWaclaw2019,Waclaw2019}, we seek a more abstract model that could serve as an ``archetype model'' for a broad class of processes that involve ``particles'' travelling down the lane that can be periodically blocked by ``obstacles'', for example RNA polymerases blocked by DNA-binding proteins, ribosomes slowed down by RNA folding, or cars stopping at traffic lights.

The model is schematically represented in Fig. \ref{fig1}A,B. Each of the $i=1,\dots,L$ lattice sites can have a particle, an obstacle, or both. We assign two occupancy variables to each site: $\tau_i$ for particles and $\sigma_i$ for obstacles. If site $i$ is occupied by a particle we set $\tau_i=1$, otherwise $\tau_i=0$. Similarly, $\sigma_i=1$ if site $i$ is occupied by an obstacle, otherwise $\sigma_i=0$. The state of the system is thus fully defined by two vectors $\bm\tau=\{\tau_i\}$ and $\bm\sigma=\{\sigma_i\}$. 

Obstacles bind and unbind with rates $k_+,k_-$. A particle jumps from $i$ to $i+1$ at rate $v$ if there is no particle at site $i+1$, and no obstacle at site $i$, otherwise the particle does not jump. We can write these rules as:
\begin{subequations}
\begin{align}
    & \sigma_i=0 \xrightarrow{k_+} 1,\\
    & \sigma_i=1 \xrightarrow{k_-} 0,\\
    & (\tau_i,\tau_{i+1})=(1,0) \xrightarrow{v(1-\sigma_i)} (0,1).
\end{align}
\end{subequations}

We consider both the closed (with periodic boundary conditions) and the open system. In the closed system, particles jump from site $i=L$ to site $i=1$, and the total number of particles is fixed and equal to $M$. In the open system, particles enter at site $i=1$ at rate $\alpha$ if the site is unoccupied. Particles leave the system from site $i=L$ at rate $\beta$ provided the site $L$ is not occupied by an obstacle. These boundary conditions can be summarised as
\begin{subequations}
\begin{align}
    (\tau_L,\tau_{1})=(1,0) \xrightarrow{v(1-\sigma_L)} (0,1) & \quad\textrm{(closed system)},\\
    \left.\begin{aligned}
        & \tau_1=0 \xrightarrow{\alpha} 1\\
        & \tau_L=1 \xrightarrow{\beta(1-\sigma_L)} 0
    \end{aligned}\right\} & \quad\textrm{(open system)}.
\end{align}
\end{subequations}
We note that the particle-obstacle interaction in this model is slightly different than in Ref.~\cite{Waclaw2019} in which a particle at site $i$ was blocked by an obstacle at site $i+1$. However, these two models are related by the particle-hole symmetry $\tau_i\leftrightarrow 1-\tau_i$ and the reversal of entrance and exit rates, $\alpha\leftrightarrow\beta$.

Using the above notation, we define the total density $\rho$ and current $J$ in the steady state as follows,
\begin{align}
    & \rho=\frac{1}{L}\sum_{i=1}^{L}\langle\tau_i\rangle,\label{eq:rhodef}\\
    & J=v\langle(1-\sigma_i)\tau_i(1-\tau_{i+1})\rangle,\label{eq:Jdef}
\end{align}
where $\langle\dots\rangle$ is taken with respect to the steady-state probability $P^{*}(C)$ to find the system in state $C=(\bm\tau;\bm\sigma)$. In the model with open boundaries, the current is related to the occupation of boundary sites as follows: $J=\alpha(1-\langle\tau_1\rangle)=\beta\langle(1-\sigma_L)\tau_L\rangle$. Since particles are conserved in the bulk, the steady-state current $J$ does not depend on position $i$ and is the same for all lattice sites. 

By definition, the current $J$ in Eq.~(\ref{eq:Jdef}) is determined by correlations between the obstacle and particle occupancy variables $\sigma_i$ and $\tau_i$ at site $i$, respectively, and the hole occupancy variable $1-\tau_{i+1}$ at site $i+1$. In the low-density limit, $J\approx v\langle\sigma_i\tau_i\rangle=u\rho$, where $u$ is the effective particle speed. The linear term in $J(\rho)$ is thus determined by particle-obstacle correlations, while non-linear terms are related to correlations involving multiple particles and obstacles.

\begin{figure}[htp]
    \centering\includegraphics[width=\columnwidth]{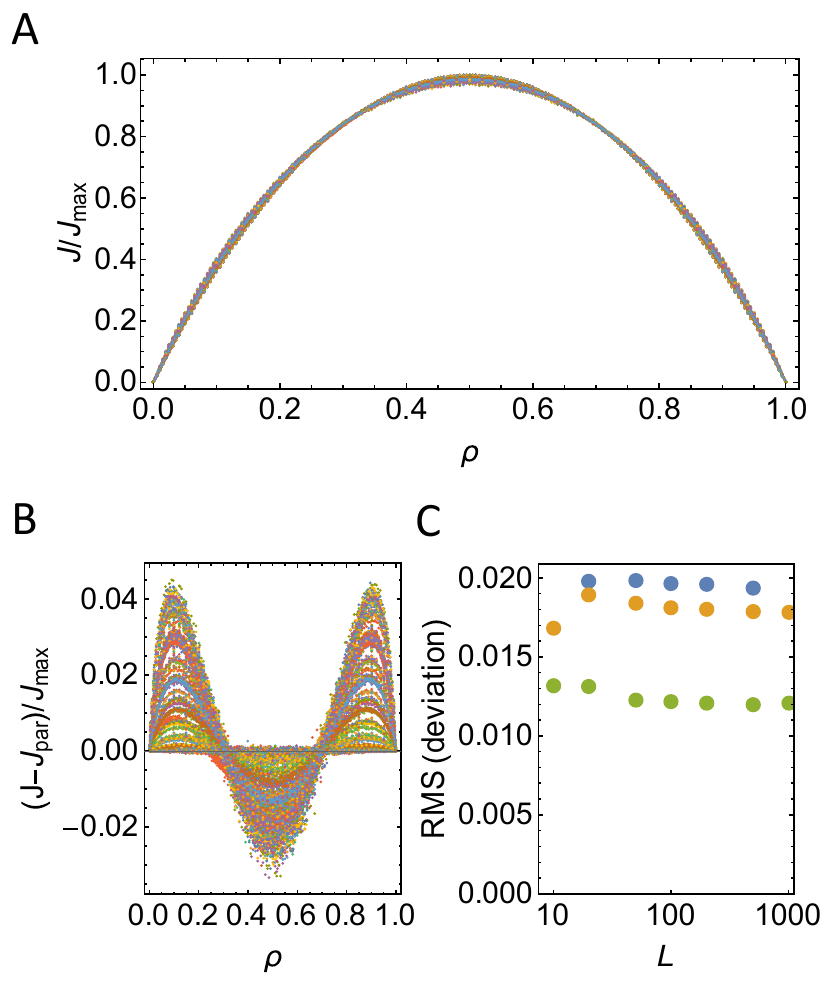}
	\caption{\label{fig2}(A) Current $J$ in the system with periodic boundaries as a function of the particle density $\rho$. Different colours represent 235 combinations of the parameters $L,k_-,k_+$: $L=10\dots 1000$ and $k_-,k_+=0.001\dots 20$. (B) Difference $(J-J_{\rm par})/J_{\rm max}$ between numerically determined $J$ and the inverted parabola $J_{\rm par}=4J_{\rm max}\rho(1-\rho)$, for the same data as in the panel (A). (C) Root-mean-square deviation $J-J_{\rm par}$ for $k_-=k_+=0.001$ (blue), $0.05$ (yellow), $0.2$ (green), for different sizes $L$. Errors (SEM) are smaller than the point size.}
\end{figure}

\section{Numerical results}
\label{simulations}

We first establish, using Monte Carlo simulations (large $L$) and exact enumeration (small $L$,) how well the current-density relation $J(\rho)$ is described by the inverted parabola from Eq.~(\ref{eq:Ju}). We set $v=1$ without loss of generality, which is equivalent to rescaling $k_+\rightarrow k_+/v$ and $k_-\rightarrow k_-/v$.

\begin{figure}[htp]
	\centering\includegraphics[width=\columnwidth]{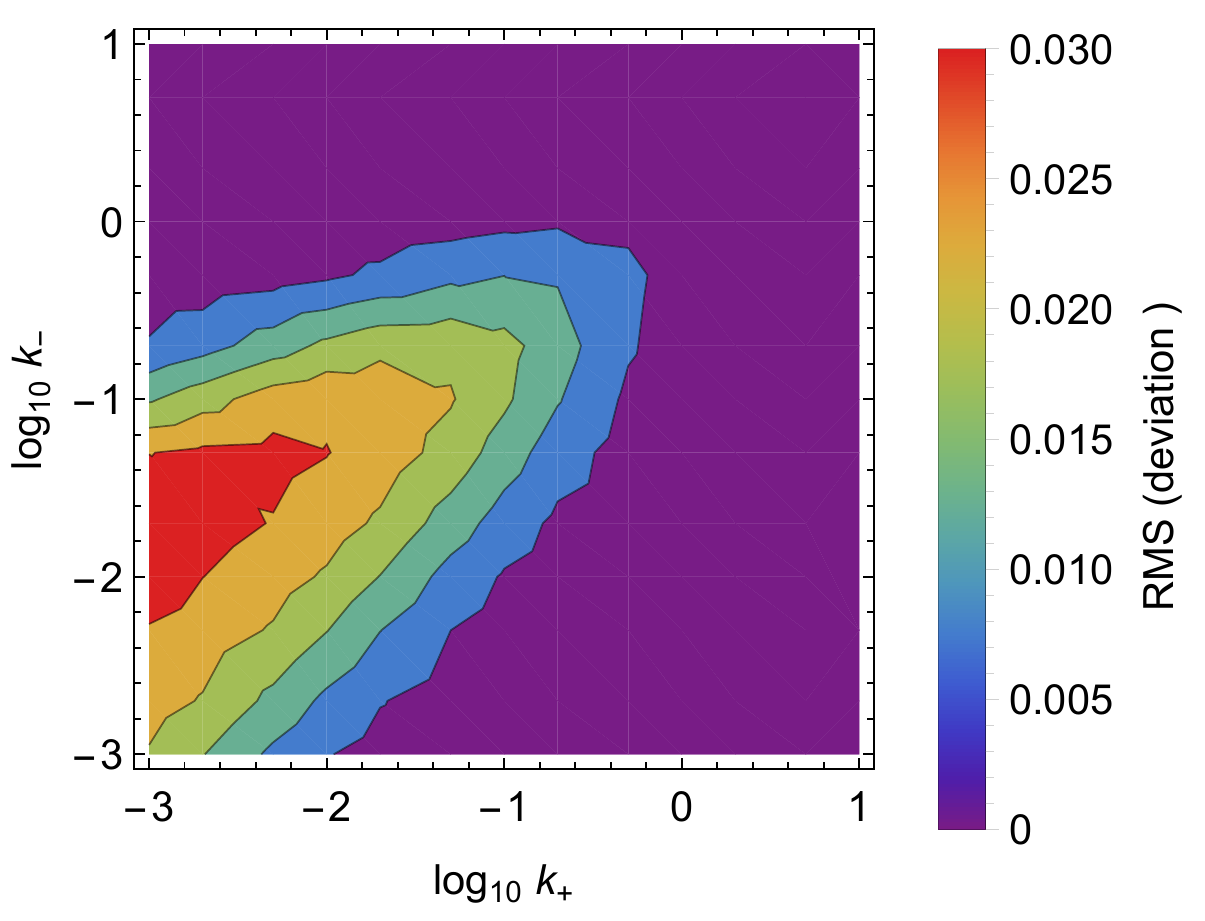}
	\caption{\label{fig3} Root-mean-square (RMS) deviation of the current $J(\rho)$ obtained in simulations from the parabola $J(\rho)=4J_{\textrm{max}}\rho(1-\rho)$, for $k_-,k_+=0.001\dots 20$ and fixed $L=1000$.}
\end{figure}

We begin with the model with periodic boundary conditions. We have simulated the model for different parameters $L,k_-,k_+$, and $M=1,\dots,L-1$. The simulation algorithm is explained in Appendix \ref{appB}. Figure \ref{fig2}A shows the scaled current $J(\rho)/J_{\textrm{max}}$ where $\rho=M/L$ and $J_{\textrm{max}}$ is the maximum value of $J$ attained at $\rho=1/2$. The shape is very well approximated by the inverted parabola $4\rho(1-\rho)$, exactly as for the standard TASEP without obstacles. Deviations from the parabola are very small (below 4\%) for all sizes $L$ (Fig.~\ref{fig2}B), but they do not seem to go away with increasing $L$, suggesting that they are not finite-size effects (Fig. \ref{fig2}C). 

As expected, the best agreement with the parabola is found in the regime $k_+,k_-\gg 1$ in which obstacles attach and detach many times between each particle jump (Fig.~\ref{fig3}, see also the mean-field theory in Section \ref{mean-field}).
Further inspection of the current dependence on $k_-$ and $k_+$ reveals that the deviations increase with decreasing $k_+$ and $k_- $. The largest deviations are found for $k_+<k_-\ll 1$. In this regime, obstacles persist for a long time; we thus expect particles to form quasi-stationary queues behind each obstacle.

\begin{figure}[htp]
	\centering\includegraphics[width=\columnwidth]{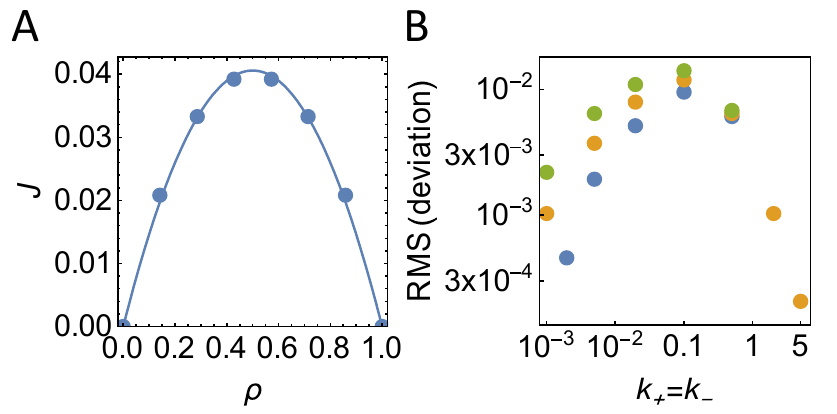}
	\caption{\label{fig4}(A) The current $J(\rho)$ obtained from exact enumeration of all states of ddTASEP (points), with a parabolic fit (line), for $L=6$ and $k_+=k_-=0.1$. (B) Root-mean-square deviation from the fitted parabola as a function of $k_+=k_-$, for three system sizes $L=5,6,7$ (blue, yellow, green).}
\end{figure}

In order to exclude the possibility that these deviations are caused by numerical inaccuracies of the simulation algorithm, we calculated $J(\rho=M/L)$ exactly for small systems for $L=5,6,7$ (see Appendix \ref{appB} for details). In the standard TASEP, the exact solution for the current for any $M$ and $L$ is given by $J=\frac{L}{L-1}\rho(1-\rho)$, i.e. the current-density relation is parabolic even for very small systems. In contrast, Figure \ref{fig4}A shows small but non-vanishing deviations from the parabola for ddTASEP. The magnitude of these deviations is the largest for intermediate binding/unbinding rates $k_+,k_-$ and decreases for very small or very large rates (Fig.~\ref{fig4}B).

\section{Mean-field theory}
\label{mean-field}

Numerical results presented in the previous section pose an interesting conundrum: the current-density relation is not exactly parabolic, but the correction never exceeds a few percent when normalised by the maximum current. The correction depends on $k_+,k_-$, being the strongest for $k_+,k_-$ much smaller than one, but decreasing again for very small rates.

In order to understand where the correction comes from, we consider the definition of the current of particles at site $i$ in Eq.~(\ref{eq:Jdef}). If we neglect correlations between particles, as well as correlations between particles and obstacles, we get the ``naive'' mean-field approximation for the current from Ref. \cite{Waclaw2019}:
\begin{equation}
	J = v(1-\rho_d)\rho(1-\rho),
	\label{eq:J-mf}
\end{equation}
where $\rho_d=k_+/(k_++k_-)$ is the equilibrium density of obstacles. We have shown in Ref.~\cite{Waclaw2019} that this approximation works very well for $k_+,k_-\gg 1$, but it breaks down for $k_+,k_-\lesssim 1$, which indicates the presence of strong correlations between variables $\tau_i$ and $\sigma_i$ from Eq. (\ref{eq:Jdef}). 

In the following sections, we will show how these correlations arise from interactions between particles and obstacles, and how including them gives a much better estimate of the current. We begin by postulating $J(\rho)\approx u\rho(1-\rho)$, which implies that $J\rightarrow u\rho$ as $\rho\rightarrow 0$. This assumption enables us to find $u$ and $J_{\rm max}=u/4$ by considering the limit $\rho\rightarrow 0$ in which particle collisions are negligible. Later we shall show how to derive the parabolic current-density relation directly from the master equation of the model, and explain the origin of higher-order corrections.

\section{Exact results for a single particle on the infinite lattice}
\label{one-particle}

We begin by presenting a simple calculation which correctly predicts the leading correction to the mean-field current in Eq.~(\ref{eq:J-mf}). 

We consider a single particle ($M=1$) on an infinite lattice ($L\to\infty$). Let $P_1(n,t)$ and $P_0(n,t)$ be the probabilities of finding the particle at site $n$ with and without an obstacle, respectively. These probabilities evolve according to the following master equation:
\begin{align}
    \frac{\dd P_1(n,t)}{\dd t} &=v\rho_d P_0(n-1,t)+k_+ P_0(n,t)\nonumber\\
    & - k_- P_1(n,t), \label{eq:dp1}\\
    \frac{\dd P_0(n,t)}{\dd t} &= v(1-\rho_d) P_0(n-1,t)+k_- P_1(n,t)\nonumber\\
    & -(k_+ + v)P_0(n,t). \label{eq:dp2}
\end{align}
In Eq.~(\ref{eq:dp1}), $\rho_d P_0(n-1,t)$ is the probability that the particle is at site $n-1$ at time $t$, and an obstacle is at site $n$. Here we have used the fact that the obstacle dynamics at site $i$ does not depend on the particle dynamics at site $i-1$, hence the product $\rho_d P_0(n-1,t)$ (see also Eq.~(\ref{eq:a_1-factor})). The second term in Eq.~(\ref{eq:dp1}), $k_+ P_0(n,t)$, accounts for an obstacle binding to site $n$ occupied currently by the particle, whereas $k_-P_1(n,t)$ corresponds to the obstacle unbinding. In the second equation, $(1-\rho_d) P_0(n-1,t)$ is the probability that the particle is at site $n-1$ at time $t$, but there is no obstacle at site $n$. The term $k_- P_1(n,t)$ represents an obstacle vanishing from the site where the particle is, and $-(k_+ + v)P_0(n,t)$ to either the particle moving away from site $n$, or an obstacle unbinding from that site.
We assume that the particle is initially at site $0$, which has no obstacle, i.e. $P_0(n,0)=0$ and $P_1(n,0)=\delta_{n,0}$, where $\delta_{n,0}$ is the Kronecker delta. Note that we did not assume $v=1$ in the equations.

\begin{figure}[htb]
	\includegraphics[width=3.4in]{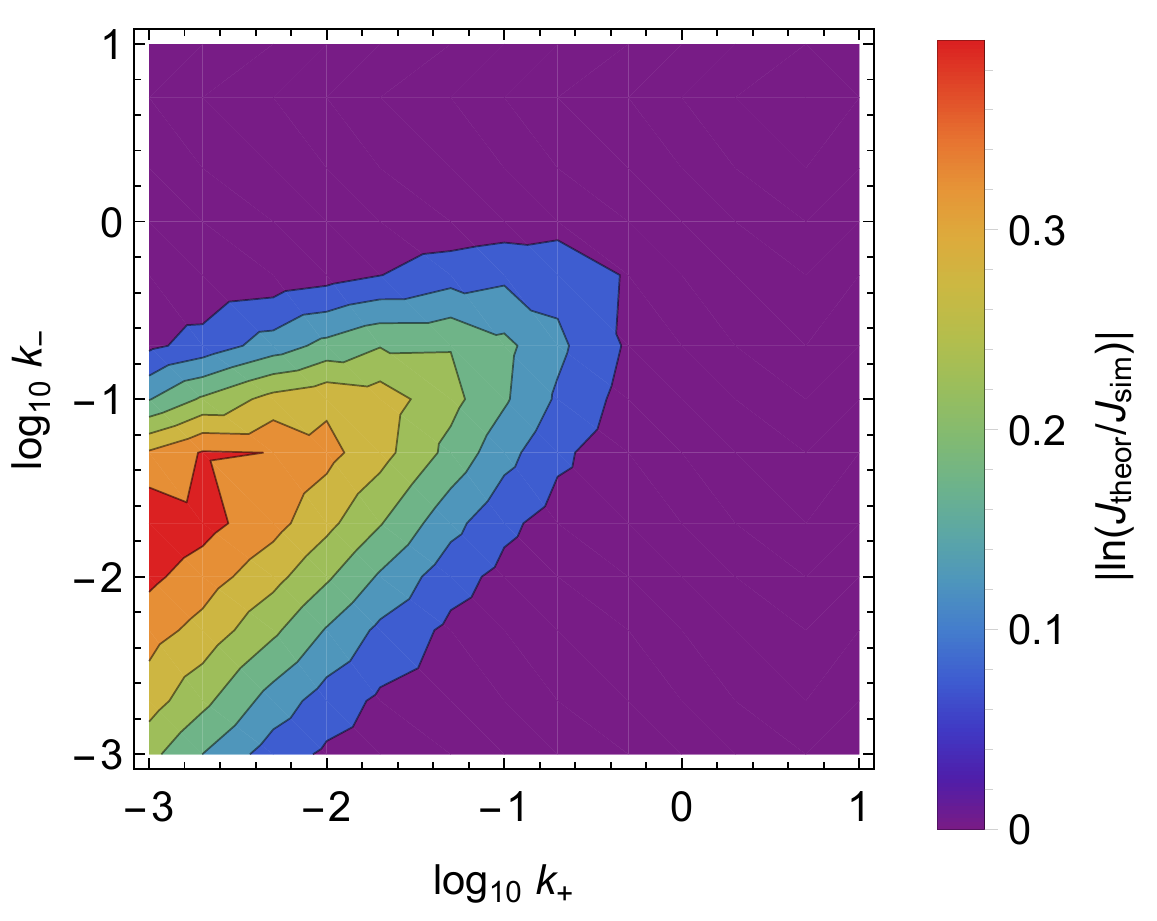}
	\caption{\label{fig5}Deviation (calculated as $\ln(J_{\rm max, theor}/J_{\rm max, sim})$) between $J_{\rm max, sim}$ from simulations and $J_{\rm max, theor}$ predicted from Eq. (\ref{eq:J1p}), for $k_-,k_+=0.001\dots 20$ and a fixed $L=1000$.}
\end{figure}

Equations (\ref{eq:dp1}, \ref{eq:dp2}) do not have a steady-state solution as the particle keeps moving through the lattice. To find the time-dependent solution, we introduce generating functions:
\begin{align}
	F_0(z,t)&=\sum_{n=0}^\infty P_0(n,t)z^n, \\
	F_1(z,t)&=\sum_{n=0}^\infty P_1(n,t)z^n.
\end{align}
The equations for $F_0(z,t)$ and $F_1(z,t)$ read
\begin{align}
    \frac{\partial F_0}{\partial t} &= \left[v(z-1)-\rho_d v z-k_+\right]F_0 + k_- F_1 \label{eq:F_0},\\
	\frac{\partial F_1}{\partial t} &= (\rho_d v z + k_+)F_0-k_-F_1\label{eq:F_1},
\end{align}
with the initial condition $F_0(z,0)=0$ and $F_1(z,0)=1$. We are interested in the speed of the particle $u$ in the long-time limit,
\begin{equation}
	u = \underset{t\rightarrow\infty}{\textrm{lim}}\frac{\partial \left<n(t)\right>}{\partial t},
\end{equation}
where $\langle n(t)\rangle$ is the mean position of the particle,
\begin{align}
	\left<n(t)\right> &= \sum_{n=0}^\infty n(P_0(n,t)+P_1(n,t))  \\
	&= \left[\frac{\partial}{\partial z} (F_0(z,t)+F_1(z,t))\right]_{z=1}.
\end{align}
In order to find $u$, we add Eqs.~(\ref{eq:F_0}) and (\ref{eq:F_1}) together and differentiate with respect to $z$ at $z=1$:
\begin{equation}
    \frac{\partial\langle n(t)\rangle}{\partial t}=\left[\frac{\partial^2}{\partial t\partial z} (F_0(z,t)+F_1(z,t))\right]_{z=1}=v F_0(1,t).
\end{equation}
Next, we note that $F_0(1,t)+F_1(1,t)=1$, which after inserting in Eq.~(\ref{eq:F_0}) gives
\begin{equation}
    \frac{\dd F_0(1,t)}{\dd t}=-(\rho_d v +k_+ + k_-)F_0(1,t)+k_-.
\end{equation}
The solution of this equation in the limit $t\rightarrow\infty$ is equal to $k_-/(\rho_d v + k_+ + k_-)$ and thus
\begin{equation}
	u = v \frac{k_-/k_+}{1+(k_-/k_+)+v/(k_++k_-)}.
	\label{eq:current-density}
\end{equation}
This must be also equal to $J/\rho$ in the limit $\rho\to 0$ and hence
\begin{equation}
	J_{\rm max} = \frac{v}{4} \frac{k_-/k_+}{1+(k_-/k_+)+v/(k_++k_-)}. \label{eq:J1p}
\end{equation}
Figure \ref{fig5} shows that this result correctly reproduces $J_{\rm max}$ measured in computer simulations to $\pm 30\%$ for $k_-,k_+$ spanning four orders of magnitude. Moreover, $J_{\rm max}$ correctly reduces to $v/4$ in the limit  $k_-/k_+\rightarrow\infty$ that corresponds to the standard TASEP without obstacles.

The fact that this simple calculation works so well means that particle-obstacle correlations are the primary factor responsible for setting the maximum current, whereas particle-particle correlations are secondary (since we have neglected them in the calculation). However, we anticipate that corrections to the parabolic shape must come from higher-order correlations involving more than one particle.

In the following section we show how to obtain Eq. (\ref{eq:J1p}) by deriving Eq. (\ref{eq:Ju}) explicitly rather than postulating it. Specifically, we will consider a system with open boundaries and expand the steady-state probability in the powers of the entry rate $\alpha$ (low-density regime) and exit rate $\beta$ (high-density regime).

\section{Exact results for the open system}
\label{series-expansion}

In this Section we use the power series method developed in Ref.~\cite{Szavits2013,Szavits2018b} to solve the steady-state master equation, first for small $\alpha$ and then for small $\beta$. For small $\alpha$, we compute the density $\rho(\alpha)$ up to the linear term in $\alpha$ and the current $J(\rho)$ up to the quadratic term in $\alpha$. We then compute the current-density relation $J(\rho)$ in the low-density regime $\rho\rightarrow 0$ by inverting $\rho(\alpha)$ into $\alpha(\rho)$ and inserting $\alpha(\rho)$ into $J(\alpha)$. We repeat this calculation for small $\beta$, which yields the current-density relation $J(\rho)$ in the high-density regime $\rho\rightarrow 1$. 

\subsection{Low-density regime}

For small $\alpha$, we expand the steady-state probability $P^{*}(C)$ as a power series in $\alpha$
\begin{equation}
    P^{*}(C)=\sum_{n=0}^{\infty}a_n(C)\alpha^n.
    \label{eq:power-series-alpha}
\end{equation}
Here, $a_n(C)$ are unknown coefficients that depend on the configuration $C$ and the model parameters other than $\alpha$. The coefficients $a_n(C)$ that can be obtained by inserting $P^{*}(C)$ into the master equation and noting that all terms with $\alpha^n$ for any $n\geq 0$ must sum to zero. The master equation is then replaced by a hierarchy of algebraic equations. These equations have the same structure as the master equation, with $P^{*}(C)$ replaced by $a_n(C)$ unless $P^{*}(C)$ is multiplied by $\alpha$, in which case it is replaced by $a_{n-1}(C)$ if $n>0$ or by zero if $n=0$. 

A crucial property of  $a_n(C)$ that simplifies the calculation is that 
\begin{equation}
    a_n(C)=0\quad\textrm{if}\; \sum_{i=1}^{L}\tau_i(C)>n.
    \label{eq:a_n_0}
\end{equation}
In other words, $a_n(C)\neq 0$ only if the number of particles in configuration $C$ is less or equal to $n$. In particular, for $n=0$, $a_0(C)\neq 0$ only if $C$ has zero particles, for $n=1$, $a_1(C)\neq 0$ only if $C$ has zero or one particle, and so on. The non-trivial condition~(\ref{eq:a_n_0}) follows from the Markov chain tree theorem \cite{Hill1966,Shubert1975,Schnakenberg1976}, which expresses the steady-state probability $P^{*}(C)$ in terms of spanning trees of a directed, weighted graph defined by the transition rate matrix (see Ref.~\cite{Szavits2018b} for more details). We further note that since $P^{*}(C)$ must sum to $1$,
\begin{equation}
    \sum_{C}a_n(C)=\delta_{n,0}.
    \label{eq:a_n-sum}
\end{equation}

Our goal is to compute $a_0(C)$ and $a_1(C)$, which in turn allows us to expand $J$ up to the term $\sim\alpha^2$ and $\rho$ up to term $\sim\alpha$.
Details of this calculation are presented below. The final result is
\begin{subequations}
\begin{align}
    \rho=\rho_{\rm LD}(\alpha) &\equiv \frac{\alpha}{u}+\mathcal{O}(\alpha^2),\label{eq:rho-alpha}\\
    J=J_{\rm LD}(\alpha) &\equiv\alpha-\frac{\alpha^2}{u}+\mathcal{O}(\alpha^3).\label{eq:J-alpha}
\end{align}
\end{subequations}
where $u$ is given by Eq.~(\ref{eq:current-density}) and the subscript LD denotes the low-density (LD) regime. From here we get $\alpha=u\rho+\mathcal{O}(\rho^2)$, which after inserting into Eq.~(\ref{eq:J-alpha}) gives the current-density relation 
\begin{equation}
    J(\rho)=u(1-\rho),
    \label{eq:J-rho-alpha}
\end{equation}
which applies to the low-density regime $\rho\rightarrow 0$.

\subsubsection{The zeroth order}

According to Eq.~(\ref{eq:a_n_0}), the coefficients $a_0(C)$ with no particles are non-zero and all other coefficients $a_0(C)$ are zero. We denote by $\bm\emptyset$ the configuration of particles $\{\tau_i\}$ where all $\tau_i=0$, and $\bm{\sigma}=\{\sigma_1,\dots,\sigma_L\}$ represents the configuration of obstacles. Let $\bm{\sigma}^{(i)}$ be a configuration derived from $\bm{\sigma}$ by replacing $\sigma_i$ for a given $i$ with $\sigma_i=1-\sigma_i$. The equation for $a_0(\bm\emptyset;\bm{\sigma_i})$ then reads
\begin{align}
    0&=\sum_{i=1}^{L}\left[k_{+}\sigma_i+k_{-}(1-\sigma_i)\right]a_0(\bm\emptyset;\bm{\sigma}^{(i)})\nonumber\\
    &-\sum_{i=1}^{L}\left[k_{-}\sigma_i+k_{+}(1-\sigma_i)\right]a_0(\bm\emptyset;\bm{\sigma}).
    \label{eq:a_0-eq}
\end{align}
This equation can be easily solved by observing that, since defects bind and unbind independently, the weight must factorise:
\begin{equation}
    a_0(\bm\emptyset;\bm{\sigma})=\prod_{i=1}^{L}g(\sigma_i).
    \label{eq:a_0}
\end{equation}
After inserting Eq.~(\ref{eq:a_0}) into (\ref{eq:a_0-eq}) we obtain that $k_{+}g(0)-k_{-}g(1)=0$. From Eq.~(\ref{eq:a_n-sum}) it follows that $g(0)+g(1)=1$. Combining this together we have
\begin{equation}
    g(0)=1-\rho_d=\frac{k_{-}}{k_{+}+k_{-}},\quad g(1)=\rho_d=\frac{k_{+}}{k_{+}+k_{-}}.
\end{equation}
We note that Eq.~(\ref{eq:a_0}) solves the original master equation when $\alpha=0$, in which case the only dynamics in the system is the binding/unbinding of obstacles.

\subsubsection{The first order}

We now turn to $a_1(C)$ whereby $C$ has at most one particle since all other coefficients $a_1(C)$ are zero. We denote by $\bm{1_i}$ the vector $\bm{\tau}$ with only one non-zero element $\tau_i=1$, i.e. a single particle is at site $i$. The equation for $a_1(\bm{1_i};\bm{\sigma})$ reads
\begin{align}
    0&=\sum_{j=1}^{L}\left[k_{+}\sigma_j+k_{-}(1-\sigma_j)\right]a_1(\bm{1_i};\bm{\sigma}^{(j)})\nonumber\\
    &-\sum_{j=1}^{L}\left[k_{-}\sigma_j+k_{+}(1-\sigma_j)\right]a_1(\bm{1_i};\bm{\sigma})\nonumber\\
    &+\begin{cases}a_0(\emptyset;\bm{\sigma}), & i=1\\ v(1-\sigma_{i-1})a_1(\bm{1_{i-1}};\bm{\sigma}), & i\neq 1\end{cases}\nonumber\\
    &-\begin{cases}v(1-\sigma_i)a_1(\bm{1_i};\bm{\sigma}), & i\neq L\\ \beta(1-\sigma_{i})a_1(\bm{1_{i}};\bm{\sigma}), & i=L\end{cases}.
    \label{eq:a_1-eq}
\end{align}
It is useful to introduce the following ``marginalised'' coefficients in which all but the specified state variables have been summed over:
\begin{subequations}
\begin{align}
    & a_1(\bm{1_i})=\sum_{\bm{\sigma}}a_1(\bm{1_i};\bm{\sigma})\\
    & a_1(\bm{1_i};\sigma_i)=\sum_{\sigma_1}\dots\sum_{\sigma_{i-1}}\sum_{\sigma_{i+1}}\dots\sum_{\sigma_L}a_1(\bm{1_i};\bm{\sigma})\\
    & a_1(\bm{1_i};\sigma_i\sigma_{i+1})=\sum_{\sigma_1}\dots\sum_{\sigma_{i-1}}\sum_{\sigma_{i+2}}\dots\sum_{\sigma_L}a_1(\bm{1_i};\bm{\sigma}).
\end{align}
\end{subequations}
We now consider the case $i=1$. By summing Eq.~(\ref{eq:a_1-eq}) over all $\{\sigma_i\}$ except for $\sigma_1$ we get a system of two equations with two unknowns, $a_1(\bm{1_1};0_1)$ and $a_1(\bm{1_1};1_1)$, 
\begin{subequations}
\begin{align}
    & -(k_+ + v)a_1(\bm{1_1};0_1)+k_- a_1(\bm{1_1};1_1)=-g(0), \label{eq:a_1-eq-10} \\
    & k_+a_1(\bm{1_1};0_1)-k_- a_1(\bm{1_1};1_1)=-g(1).
    \label{eq:a_1-eq-11}
\end{align}
\end{subequations}
The solution is:
\begin{equation}
    a_1(\bm{1_1};0_1)=\frac{1}{v},\quad a_1(\bm{1_1};1_1)=\frac{k_+}{k_-v}+\frac{g(1)}{k_-},
    \label{eq:a_1-solution-1}
\end{equation}
which gives $a_1(\bm{1_1})=(1/v+g(1)g(0)/k_-)/g(0)$. If we repeat the procedure for $i=2,\dots,L-1$, we get the following equations for $a_1(\bm{1_i};0_i)$ and $a_1(\bm{1_i};1_i)$,
\begin{subequations}
\begin{align}
    &-(k_+ + v)a_1(\bm{1_i};0_i)+k_- a_1(\bm{1_i};1_i) =\nonumber\\
    &\quad-va_1(\bm{1_{i-1}};0_{i-1}0_i), \label{eq:a_1-eq-i0} \\
    &k_+a_1(\bm{1_i};0_i)-k_- a_1(\bm{1_i};1_i) =\nonumber\\
    &\quad-va_1(\bm{1_{i-1}};0_{i-1}1_i).
    \label{eq:a_1-eq-i1}
\end{align}
\end{subequations}
Since an obstacle at site $i$ does not affect the particle at site $i-1$, it follows that
\begin{equation}
a_1(\bm{1_{i-1}};0_{i-1}\sigma_i)=a_1(\bm{1_{i-1}};0_{i-1})g(\sigma_i).
\label{eq:a_1-factor}
\end{equation}
The system of equations for $a_1(\bm{1_i};\sigma_i)$ can now be solved recursively, and the final expressions for $a_1(\bm{1_i};0_i)$ and $a_1(\bm{1_i};1_i)$ are the same as in Eq.~(\ref{eq:a_1-solution-1}), except for $i=L$ for which $v$ is replaced by $\beta$. The coefficients $a_1(\bm{1_i})$ are thus given by
\begin{subequations}
\begin{align}
    & a_1(\bm{1_i})=\frac{1}{u},\quad i=1,\dots,L-1,\\
    & a_1(\bm{1_L})=\frac{1}{\beta}\left(1+\frac{k_+}{k_-}\right)+\frac{k_+}{k_-(k_+ + k_-)},
\end{align}
\end{subequations}
where $u=g(0)/(1/v+g(1)g(0)/k_-)$ is the same as in  Eq.~(\ref{eq:current-density}).

Now that we know the probability of configurations with zero or one particle, we can write down the expression for the total density $\rho$ of particles,
\begin{align}
    \rho &=\frac{1}{L}\sum_{i=1}^{L}\alpha a_1(\bm{1_i})+\mathcal{O}(\alpha^2)\nonumber\\
    &=\alpha\left[\frac{1}{u}+\mathcal{O}(1/L)\right]+\mathcal{O}(\alpha^2). \label{eq:rho-alpha1}
\end{align}
and the current $J$,
\begin{align}
J &= \alpha +\alpha^2\left(a_1(\emptyset)+\sum_{i=2}^{L}a_1(\bm{1_i})\right)+\mathcal{O}(\alpha^3)\nonumber\\ 
&=\alpha-\frac{\alpha^2}{u}+\mathcal{O}(\alpha^3), \label{eq:J-alpha2}
\end{align}
where in the last expression we have used $a_1(\bm\emptyset)=-\sum_{i=1}^{L}c(\bm{1_i})$, which follows from Eq.~(\ref{eq:a_n-sum}). 

We note that there is a contribution to $J$ of order $O(\rho^2)$ that comes from the quadratic term in the series expansion of $\rho(\alpha)=\alpha/u+\rho_2 \alpha^2+\mathcal{O}(\alpha^3)$. After inverting $\rho(\alpha)$ we get $\alpha(\rho)=u\rho-\rho_{2}u^3\rho^2+\mathcal{O}(\rho^3)$, which, when inserted into $J(\alpha)$, gives an extra contribution $-\rho_{2}u^2\rho^2$ to $J(\rho)$. Unfortunately, we could not find an expression for $\rho_2$ in terms of $k_+,k_-$ as the second order turned out to be a difficult problem.

\begin{figure*}[htb]
    \centering
    \includegraphics[width=\textwidth]{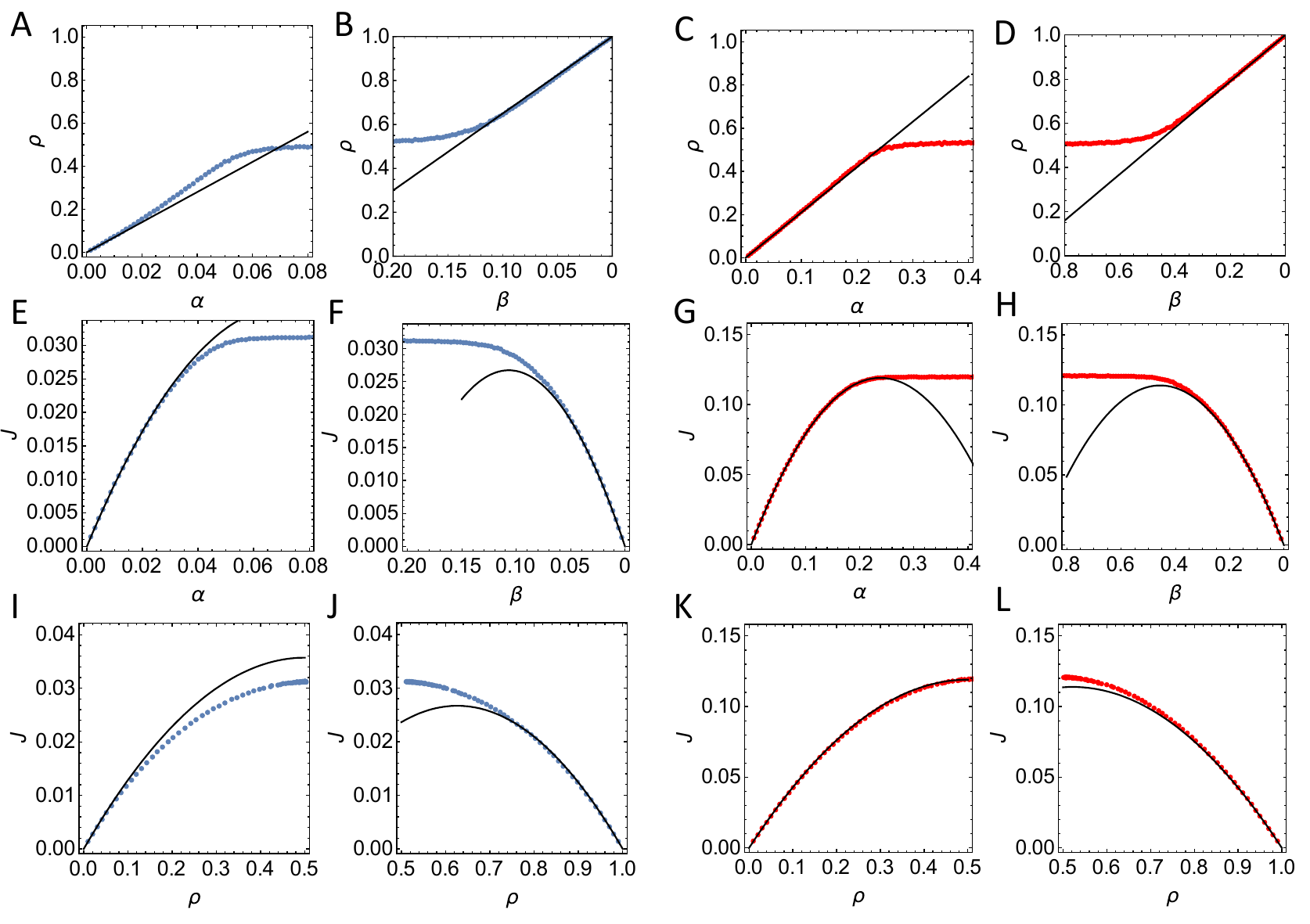}
    \caption{\label{fig6} Particle density $\rho$ and particle current $J$ in the system with open boundaries. Left-hand panels (A, B, E, F, I and J): $k_-=k_+=0.1$. Right-hand panels (C, D, G, H, K and L): $k_-=k_+=5$. Other parameters common to all figures are $v=1,L=100$. 
    Panels (A, C, E, G, I and K) correspond to small-$\alpha$ expansion (fixed $\beta=1$), panels (B, D, F, H, J and L) to small-$\beta$ expansion (fixed $\alpha=1$). Points represent Monte-Carlo simulations, lines are analytic formulas: Eq. (\ref{eq:rho-alpha}) for panels A and C, Eq. (\ref{eq:rho-beta}) for panels B and D, Eq. (\ref{eq:J-alpha}) for panels E and G, and Eq. (\ref{eq:J-beta}) for panels F and H. $J(\rho)$ in panels I,J,K and L has been obtained by inverting the equations for $\rho(\alpha),\rho(\beta)$ for $\alpha=\alpha(\rho),\beta=\beta(\rho)$ and inserting them into the equations for $J(\alpha),J(\beta)$.}
\end{figure*}

In Fig.~\ref{fig6} we compare the predicted $\rho(\alpha)$, $J(\alpha)$ and $J(\rho)$ with the results from numerical simulations for $k_+=k_-=0.1$ and $5$, while $\beta=1$ is fixed. The density $\rho(\alpha)$ grows linearly for small $\alpha$ as predicted by Eq.~(\ref{eq:rho-alpha}) and approaches $1/2$ for large $\alpha$ (Fig.~\ref{fig6}A and C). The transition from $\rho=\alpha/u$ to $\rho=1/2$ is sharp when $k_-$ and $k_+$ are large (Fig.~\ref{fig6}C), but smooths out for small $k_-$ and $k_+$ (Fig.~\ref{fig6}A). Similarly, the deviation from $u\rho(1-\rho)$ increases as $k_-$ and $k_+$ become smaller (Fig. \ref{fig6}I). These results are expected since decreasing $k_-$ and $k_+$ creates long-lived obstacles leading to particle entrapment not accounted for in the first order. 

\subsection{High-density regime}

In this Section we expand the steady-state probability $P^{*}(C)$ in $\beta$,  
\begin{equation}
    P^{*}(C)=\sum_{n=0}^{\infty}b_n(C)\beta^n,
    \label{eq:power-series-beta}
\end{equation}
where $b_n(C)$ are unknown coefficients. Our goal is to find $b_0(C)$ and $b_1(C)$, which will allow us to expand $\rho(\beta)$ up to the linear order and $J(\beta)$ up to the quadratic order and thus give us the expression for $J(\rho)$ that is valid close to $\rho=1$. We leave details of this calculation for later and present here the final result, which is:
\begin{equation}
    \rho=\rho_{\rm HD}(\beta)\equiv 1-\frac{k_-}{(k_++k_-)u}\beta+\mathcal{O}(\beta^2)
    \label{eq:rho-beta}
\end{equation}
and 
\begin{widetext}
\begin{equation}
    J=J_{\rm HD}(\beta)\equiv\frac{k_-}{k_++k_-}\beta-\frac{k_-\left[2(k_++k_-)^3+(k_++k_-)(2k_-+5k_+)v+2k_+v^2\right]}{v(k_++k_-)^2\left[2(k_++k_-)^2+(2k_-+k_+)v\right]}\beta^2+\mathcal{O}(\beta^3).
    \label{eq:J-beta}
\end{equation}
\end{widetext}
where the subscript HD denotes the high-density regime. From here, we obtain $J(\rho)$ by inserting $\beta=(1-\rho)u/g(0)$ into Eq.~(\ref{eq:J-beta}). 

In Fig.~\ref{fig6} we compare the predicted $\rho(\beta)$, $J(\beta)$ and $J(\rho)$ with the results from numerical simulations for $k_+=k_-=0.1$ and $5$, while $\alpha=1$ is held fixed. Interestingly, the density $\rho(\beta)$ maintains a linear profile even for small values of $k_-$ and $k_+$, suggesting that the quadratic term in the expansion of $\rho(\beta)$ is either zero or is very small (Fig.~\ref{fig6}B and D). We also note that the expressions (\ref{eq:J-alpha}) and (\ref{eq:J-beta}) for the current in the low/high density regime are not symmetric with respect exchanging particles with holes ($\tau_i\leftrightarrow 1-\tau_i$) and switching the entry and exit rates  ($\alpha\leftrightarrow\beta$). We will comment on this in the discussion. 

In the rest of this Section we derive Eqs.~(\ref{eq:rho-beta}) and (\ref{eq:J-beta}). To this end, the following two relations will be useful. First, $b_n(C)$ is zero if the number of empty sites (``holes'') in $C$ is larger than $n$,
\begin{equation}
    b_n(C)=0\quad\textrm{if}\; \sum_{i=1}^{L}[1-\tau_i(C)]>n,
    \label{eq:b_n_0}
\end{equation}
which follows from the matrix tree theorem discussed before. Second, the sum of $b_n(C)$ over all $C$ is equal to $1$ for $n=0$ and is zero otherwise,
\begin{equation}
    \sum_{C}b_n(C)=\delta_{n,0}.
    \label{eq:b_n-sum}
\end{equation}

\subsubsection{The zeroth order}

The zeroth order is equivalent of setting $\beta=0$ in the master equation, which has the following solution
\begin{equation}
    b_0(C)=\prod_{i=1}^{L}\left[\delta_{\tau_i,1}g(\sigma_i)\right],
\end{equation}
in which all sites are occupied by particles. We refer to this state as $b_0(\bm{1};\bm{\sigma}$), where $\bm{1}=\{\tau_1,\dots,\tau_L\}$ in which all $\tau_i=1$ and $\bm{\sigma}=\{\sigma_1,\dots,\sigma_L\}$.

\subsubsection{The first order}

We denote by $\bm{\emptyset_i}$ a configuration of particles with a hole at site $i$, i.e. $\tau_j=1$ for $j\neq i$ and $\tau_i=0$. According to Eq.~(\ref{eq:b_n_0}), the coefficients  $b_1(\bm{\emptyset_i};\bm{\sigma})$ and $b_1(\bm{1};\bm{\sigma})$ are non-zero and all other coefficients $b_1(C)$ are zero. As before, $\bm{\sigma}$ denotes a configuration of obstacles and $\bm{\sigma}^{(j)}$ is obtained from $\bm{\sigma}$ by replacing $\sigma_j$ with $1-\sigma_j$. The equation for $b_1(\bm{\emptyset_i};\bm{\sigma})$ reads
\begin{align}
    0&=\sum_{j=1}^{L}\left[k_{+}\sigma_j+k_{-}(1-\sigma_j)\right]b_1(\bm{\emptyset_i};\bm{\sigma}^{(j)})\nonumber\\
    &-\sum_{j=1}^{L}\left[k_{-}\sigma_j+k_{+}(1-\sigma_j)\right]b_1(\bm{\emptyset_i};\bm{\sigma})\nonumber\\
    &+\begin{cases}b_0(\bm{1};\bm{\sigma}), & i=L\\ v(1-\sigma_{i})a_1(\bm{\emptyset_{i+1}};\bm{\sigma}), & i\neq L\end{cases}\nonumber\\
    &-\begin{cases}v(1-\sigma_{i-1})a_1(\bm{\emptyset_i};\bm{\sigma}), & i=2,\dots,L\\ \alpha a_1(\bm{\emptyset_{1}};\bm{\sigma}), & i=1\end{cases}.
    \label{eq:b_1-eq}
\end{align}
As before, we introduce ``marginalised'' coefficients obtained by summing $b_1(\bm{\tau};\bm{\sigma})$ over a subset of obstacle variables $\bm{\sigma}$. In particular, we define
\begin{subequations}
\begin{align}
    & b_1(\bm{\tau})=\sum_{\sigma_1}\dots\sum_{\sigma_L}b_1(\bm{\tau};\bm{\sigma}),\\
    & b_1(\bm{\tau};\sigma_{j})=\sum_{\sigma_1}\dots\sum_{\sigma_{j-1}}\sum_{\sigma_{j+1}}\dots\sum_{\sigma_L}b_1(\bm{\tau};\bm{\sigma}),\\
    & b_1(\bm{\tau};\sigma_{j}\sigma_{k})=\sum_{\bm{\sigma}\setminus\{\sigma_j,\sigma_k\}}b_1(\bm{\tau};\bm{\sigma}),
\end{align}
\end{subequations}
where $\bm{\sigma}\setminus\{\sigma_j,\sigma_k\}$ denotes $\bm{\sigma}$ without $\sigma_j$ and $\sigma_k$. 
We first find the coefficients $b_1(\bm{\emptyset_i})$ and $b_1(\bm{1})$ which we need to compute the next term in the series expansion of the density $\rho(\beta)$. We start from $i=L$ and write the equations for $b_1(\bm{\emptyset_L};\sigma_{L-1})$,
\begin{subequations}
\begin{align}
& k_- b_1(\bm{\emptyset_L};1_{L-1})-(k_{+}+v) b_1(\bm{\emptyset_L};0_{L-1})\nonumber\\
&\quad =-g(0)g(0),\\
& k_+ b_1(\bm{\emptyset_L};0_{L-1})-k_- b_1(\bm{\emptyset_L};1_{L-1})=-g(1)g(0).
\end{align}
\end{subequations}
By solving these equations we get
\begin{subequations}
\begin{align}
    & b_1(\bm{\emptyset_L};0_{L-1})=\frac{g(0)}{v},\label{eq:b_1-L0}\\
    & b_1(\bm{\emptyset_L};1_{L-1})=\frac{g(1)}{v}+\frac{g(1)g(0)}{k_-},\label{eq:b_1-L1}
\end{align}
\end{subequations}
and thus $b_1(\bm{\emptyset_L})=1/v+g(1)g(0)/k_-$. Next, we solve the equations for $i=2,\dots,L-1$,
\begin{subequations}
\begin{align}
& k_- b_1(\bm{\emptyset_i};1_{i-1})-(k_{+}+v) b_1(\bm{\emptyset_i};0_{i-1})\nonumber\\
&\quad =-v b_1(\bm{\emptyset_{i+1}};0_{i-1}0_i),\\
& k_+ b_1(\bm{\emptyset_i};0_{i-1})-k_- b_1(\bm{\emptyset_i};1_{i-1})\nonumber\\
&\quad =-v b_1(\bm{\emptyset_{i+1}};1_{i-1}0_i).
\end{align}
\end{subequations}
Using $b_1(\bm{\emptyset_{i+1}};\sigma_{i-1}0_i)=g(\sigma_{i-1})b_1(\bm{\emptyset_{i+1}};0_i)$, we get the same result as in Eqs.~(\ref{eq:b_1-L0}) and (\ref{eq:b_1-L1}),
\begin{subequations}
\begin{align}
    & b_1(\bm{\emptyset_i};0_{i-1})=\frac{g(0)}{v}\label{eq:b_1-i0}\\
    & b_1(\bm{\emptyset_i};1_{i-1})=\frac{g(1)}{v}+\frac{g(1)g(0)}{k_-}\label{eq:b_1-i1}.
\end{align}
\end{subequations}
which together yields $b_1(\bm{\emptyset_i})=1/v+g(1)g(0)/k_-$ for $i=2,\dots,L$. Finally, we solve the equations for $b_1(\bm{\emptyset_1};\sigma_1)$:
\begin{subequations}
\begin{align}
& k_- b_1(\bm{\emptyset_1};1_1)-(k_{+}+\alpha) b_1(\bm{\emptyset_1};0_1)=0,\\
& k_+ b_1(\bm{\emptyset_1};0_1)-k_- b_1(\bm{\emptyset_1};1_1)=0,
\end{align}
\end{subequations}
which gives $b_1(\bm{\emptyset_1})=g(0)/\alpha$. We can now compute the density $\rho(\beta)$ which reads
\begin{align}
    \rho(\beta) &= 1+\frac{\beta}{L}\left[(L-1)\sum_{i=1}^{L}b_1(\bm{\emptyset_i})+Lb_1(\bm{1})\right]+O(\beta^2)\nonumber\\
    &= 1-\beta\left[\frac{g(0)}{u}+\mathcal{O}(1/L)\right]+O(\beta^2),
\end{align}
where we have used Eq.~(\ref{eq:b_n-sum}) to eliminate $b_1(\bm{1})$.

The next step is to compute $J(\beta)$ from
\begin{align}
    J(\beta) &= g(0)\beta+\left[\sum_{i=1}^{L-1}b_1(\bm{\emptyset_i};0_L)+b_1(\bm{1};0_L)\right]\beta^2\nonumber\\
    &\quad +O(\beta).
    \label{eq:J-beta-1-def}
\end{align}
We shall skip the full derivation of $J(\beta)$ for brevity, and only outline its main steps. We first find $b_1(\bm{1};0_L)$ from Eq.~(\ref{eq:b_1-eq}) by summing over all $\bm{\sigma}$ except at site $L$ at which $\sigma_L=0$, which leads to
\begin{equation}
    b_1(\bm{1};0_L)=g(0)b_1(\bm{1})-\frac{\alpha}{k_++k_-}b_1(\bm{\emptyset_1};1_L).
    \label{eq:b_1-1;0_L}
\end{equation}
Here $b_1(\bm{\emptyset_1};1_L)$ is unknown, but that is not a problem as it will cancel later. We can eliminate $b_1(\bm{1})$ using  $b_1(\bm{1})=-\sum_{i}b_1(\bm{\emptyset_i})$ which follows from Eq.~(\ref{eq:b_n-sum}). 

Next, we consider the equation for $b_1(\bm{\emptyset_i};0_L)$ obtained by summing Eq.~(\ref{eq:b_1-eq}) over all $\bm{\sigma}$ except at site $L$ for which we set $\sigma_L=0$. We get one such equation for each $i=1,\dots,L$. We then sum all these equations together, which gives
\begin{align}
    0 &= k_-\sum_{i=1}^{L}b_1(\bm{\emptyset_i};1_L)-k_+\sum_{i=1}^{L}b_1(\bm{\emptyset_i};0_L)\nonumber\\
    &\quad +g(0)-\alpha b_1(\bm{\emptyset_1};0_L),
    \label{eq:b_1-iL-sum}
\end{align}
where the last two terms can be replaced by $\alpha b_1(\bm{\emptyset_1};1_L)$ since $b_1(\bm{\emptyset_1})=g(0)/\alpha$. After inserting Eqs.~(\ref{eq:b_1-1;0_L}) and (\ref{eq:b_1-iL-sum}) into Eq.~(\ref{eq:J-beta-1-def}), we observe that many terms cancel and we are left with a simple expression $J(\beta)=g(0)\beta-b_1(\bm{\emptyset_L};0_L)\beta^2+\mathcal{O}(\beta^3)$. 

In order to find $b_1(\bm{\emptyset_L};0_L)$ we solve the equations for $b_1(\bm{\emptyset_L};\sigma_{L-1}\sigma_L)$ obtained from Eq.~(\ref{eq:b_1-eq}) by summing over all $\bm{\sigma}$ except the last two sites. After obtaining $b_1(\bm{\emptyset_L};0_{L-1}0_L)$ and $b_1(\bm{\emptyset_L};1_{L-1}0_L)$, we add them together to get $b_1(\bm{\emptyset_L};0_L)$, which we insert into the expression for $J(\beta)$ to get Eq.~(\ref{eq:J-beta}).

We note that, similarly to the small-$\alpha$ expansion, there is a contribution to $J(\rho)$ of order $\mathcal{O}(\rho^2)$ coming from the quadratic term in the series expansion of $\rho(\beta)$ which in turn comes from the second order of the series expansion in Eq.~(\ref{eq:power-series-beta}). We did not manage to find an expression for this quadratic term.

\section{Discussion}

Our main goal in this work was to better understand the quasi-parabolic current-density relation observed numerically in the TASEP with dynamic obstacles. Here we list the most important results.

{\bf The current-density relation is nearly a parabola.} According to Fig.~\ref{fig2}A, the current $J(\rho)/J_{\rm max}$ is approximately equal to $4\rho(1-\rho)$, where $J_{\rm max}$ is the maximum current in the system. This approximation is valid within the percentage error less than 5\% over the whole range of binding and unbinding rates explored in numerical simulations spanning four orders of magnitude from $k_+,k_- =10^{-3}$ to $10^1$.

{\bf Single particle-obstacle correlations explain most of the observed reduction in $J_{\rm max}$.} Since $J(\rho)\approx 4J_{\rm max}\rho(1-\rho)$, the maximum current $J_{\rm max}\approx u/4$, where $u=\lim_{\rho\rightarrow 0}J(\rho)/\rho$ is the effective particle speed in the low-density limit. As discussed in Section \ref{one-particle}, $u$ is determined by single particle-obstacle correlations, which can be computed exactly. The expression for $u$ is given by Eq.~(\ref{eq:current-density}).

{\bf The current-density relation is symmetric with respect to $\rho\leftrightarrow 1-\rho$}. Numerical results presented in Fig.~\ref{fig2}A suggest that the current-density relation obeys $J(\rho)=J(1-\rho)$. According to Fig.~\ref{fig2}B, this symmetry is preserved even after the parabolic part is subtracted from $J(\rho)$. We note that the model is not symmetric with respect to the particle-hole inversion $\tau_i\leftrightarrow 1-\tau_i$. Instead, $\tau_i\leftrightarrow 1-\tau_i$ leads to a different model in which the particle is blocked by an obstacle in front of it rather than at the same site. This model was previously studied in Ref. \cite{Waclaw2019} and exhibits the same quasi-parabolic $J(\rho)$. The fact that the two models have the same quasi-parabolic $J(\rho)$ suggests a deeper symmetry that is not immediately obvious at the level of microscopic dynamics of the models.

{\bf Corrections to the parabola show complex dependence on the density.} If we assume that $J(\rho)=J(1-\rho)$ (which is supported by our numerical results), then $J(\rho)$ must also be a function of $\rho(1-\rho)$ (see Appendix \ref{appC}). Expanding $J(\rho)$ in $\rho(1-\rho)$ gives $J(\rho)=u\rho(1-\rho)+\sum_{n=2}^{\infty}a_n\rho^n(1-\rho)^n$, where $u$ is given by Eq.~(\ref{eq:current-density}) and $a_n$ are unknown coefficients. We note that $J_{\rm par}(\rho)=4J_{\rm max}\rho(1-\rho)$ in Fig.~\ref{fig2}, where $J_{\rm max}$ is obtained by fitting an inverse parabola to $J(\rho)$ and is different than $J_{\rm max}$ given by Eq.~(\ref{eq:J1p}). Thus $J(\rho)-J_{\rm par}(\rho)$ equals $(u-4J_{\rm max})\rho(1-\rho)+\sum_{n=2}^{\infty}a_n\rho^{n}(1-\rho)^n$ and the fact that it changes the sign suggests that at least $a_2\neq 0$. This means that the exact $J(\rho)$ has a rather complicated dependence on $\rho$, although it remains unclear why the coefficients $a_n$ are small.

{\bf The current and density are asymmetric with respect to $\tau_i\leftrightarrow 1-\tau_i$ and $\alpha\leftrightarrow\beta$}. We have noted that Eqs. (\ref{eq:J-alpha}) and (\ref{eq:J-beta}) for the current in the low and high density regime are not symmetric upon replacing $\alpha\rightarrow\beta$, i.e. $J_{\rm LD}(\alpha)\neq J_{\rm HD}(\alpha)$. Similarly, Eqs. (\ref{eq:rho-alpha}) and (\ref{eq:rho-beta}) for the density $\rho_{LD}(\alpha)$ and $\rho_{\rm HD}(\beta)$ do not obey $\rho_{\rm LD}(\alpha)=1-\rho_{\rm HD}(\alpha)$. Given that the microscopic dynamics does not obey the particle-hole symmetry, these results are not surprising. However, it is surprising that the particle-hole symmetry is restored when the current is expressed as a function of the density.

{\bf Obstacle dynamics independent from particle dynamics is crucial for the observed symmetry $J(\rho)=J(1-\rho)$}. In \cite{Waclaw2019} we considered a model in which obstacles interacted with particles by exclusion. That model did not have the $\rho\to 1-\rho$ symmetry in $J(\rho)$. This suggests that the independence of the dynamics of obstacles from that of particles (but not vice versa) is crucial for the symmetric relation.

\section{Conclusions}

To summarise, we have used numerical simulations and analytic calculations to study the current-density relation in the TASEP with dynamic obstacles. We have shown that the current-density relation is not a perfect parabola, but that corrections to the parabola remain small over a large range of binding and unbinding rates. We have derived an analytic expression for the prefactor of the parabolic part of $J(\rho)$ that is in excellent agreement with the results of numerical simulations.

Interestingly, our numerical results indicate that the current-density relation $J(\rho)$ obeys $J(\rho)=J(1-\rho)$ in spite of the fact that the microscopic dynamics is not symmetric with respect to the particle-hole inversion. The origin of this symmetry remains unclear.

\begin{acknowledgments}
J.S.N. was supported by the Leverhulme Trust Early Career Fellowship under Grant No. ECF-2016-768.
\end{acknowledgments}

\appendix
\section{Monte Carlo simulations}
\label{appA}

We use a kinetic Monte Carlo method analogous to Gillespie's algorithm \cite{gillespie_exact_1977}. In each time step, we calculate the total rate of all processes that can occur: particles moving by one site (if not blocked by an obstacle or a particle), particles jumping in/out of the system at sites $i=1,L$ (in the open boundary version), obstacles binding/unbinding. Since all particles and obstacles have the same jumping and binding/unbinding rates, to speed up the calculation of the total rate we keep track of the number of particles that are allowed to move, the number of sites devoid of obstacles, and the number of obstacles. Next, we select one of the events to occur (particle moving, obstacle binding/unbinding) with probability proportional to the total probability of all events in that class. We then select a specific particle/obstacle to move/bind/unbind from an array of stored positions. Finally, we update the state of the particle/obstacle and all their associated variables. The algorithm is very fast and has an approximately constant execution time per time step. 
To determine the current for a given density of particles, we start from a random configuration of particles, and no obstacles. We first do an equilibration run, waiting for $10^6\dots 10^7$ particle hops to occur. We then run the algorithm for another $10^6\dots 10^7$ particle hops and calculate the current as the number of hops divided by the elapsed time. We repeat this process until the value of the current stabilises within 0.1\% of the value from the previous run.

We run our simulations on the School of Physics and Astronomy compute cluster. We processed the data using a Wolfram Mathematica script. The simulation code (C++) and the Mathematica script are available at 
\url{https://www2.ph.ed.ac.uk/~bwaclaw/ddTASEP/data_and_code.zip}

\section{Exact enumeration}
\label{appB}

We use Wolfram Mathematica to generate a set of algebraic equations for steady-state probabilities $P(\tau,\sigma)$ of all possible configurations of particles $\{\tau\}$ and defects $\{\sigma\}$. For given $L,M$, we go through all $2^L$ particle configurations, accepting those in which the number of particles equals $M$. For each such configuration, we go through all $2^L$ configurations of defects and use the master equation to generate a linear equation for $P(\tau,\sigma)$ for each ${\tau,\sigma}$.

The complexity of this algorithm is approximately $O(2^{2L})$ for the equation generation part, and $O(2^{4L})$ for the equation solving part. The memory usage is also approximately $O(2^{4L})$. The algorithm is thus suitable only for very small systems. We have used it for $L\leq 7$. The implementation of the algorithm as a Mathematica script is available at \url{https://www2.ph.ed.ac.uk/~bwaclaw/ddTASEP/data_and_code.zip}

\section{\texorpdfstring{$J(\rho)$ is a function of $\rho(1-\rho)$}{J(rho) is a function of rho(1-rho)}}
\label{appC}

Here we show that if $J(\rho)=J(1-\rho)$ is satisfied for $0\leq \rho \leq 1$ and $J(0)=J(1)=0$, then $J(\rho)$ is a function of $\rho(1-\rho)$. 

To arrive at this result we expand $J(\rho)$,
\begin{equation}
    J(\rho)=\sum_{n=1}^{\infty}\frac{J^{(n)}(0)}{n!}\rho^n.
\end{equation}
Assuming $J(\rho)=J(1-\rho)$, we can also write
\begin{equation}
    J(1-\rho)=\sum_{n=1}^{\infty}\frac{J^{(n)}(0)}{n!}(1-\rho)^n.
\end{equation}
Combining these two expression we can write
\begin{equation}
    J(\rho)=\frac{J(\rho)+J(1-\rho)}{2}=\sum_{n=1}^{\infty}\frac{J^{(n)}(0)}{n!}S_n,
\end{equation}
where we have introduced $S_n=[\rho^n+(1-\rho)^n]/2$. We can check that $S_n$ satisfies the following recurrence relation
\begin{equation}
    S_{n+1}=S_n-\rho(1-\rho)S_{n-1}, 
\end{equation}
where $S_0=1$ and $S_1=1/2$. From here we conclude that $S_n$ and hence $J(\rho)$ are both functions of $\rho(1-\rho)$.

\bibliography{references}

\end{document}